\documentclass[aps,prl,superscriptaddress,tightenlines,footinbib,longbibliography,twocolumn]{revtex4-2}
\usepackage{graphicx}
\usepackage[colorlinks,linkcolor=blue,anchorcolor=blue,citecolor=blue,urlcolor=blue]{hyperref}
\usepackage{amsmath}
\usepackage{color}
\usepackage{soul}
\usepackage{ulem}

\begin{document}
\title{Observation of $\mathcal{PT}$-Symmetric Quantum Coherence in a Single Ion System}


	
\author{Wei-Chen Wang}
\thanks{These authors contributed equally to this work.}
\affiliation{Department of Physics, College of Liberal Arts and Sciences, National University of Defense Technology, Changsha 410073, China}
\affiliation{Interdisciplinary Center for Quantum Information, National University of Defense Technology, Changsha 410073, China}

\author{Yan-Li Zhou}
\thanks{These authors contributed equally to this work.}
\affiliation{Department of Physics, College of Liberal Arts and Sciences, National University of Defense Technology, Changsha 410073, China}
\affiliation{Interdisciplinary Center for Quantum Information, National University of Defense Technology, Changsha 410073, China}

\author{Hui-Lai Zhang}
\affiliation{Key Laboratory of Low-Dimensional Quantum Structures and
	Quantum Control of Ministry of Education, and Department of Physics,
	Hunan Normal University, Changsha 410081, China}

\author{Jie Zhang}
\affiliation{Department of Physics, College of Liberal Arts and Sciences, National University of Defense Technology, Changsha 410073, China}
\affiliation{Interdisciplinary Center for Quantum Information, National University of Defense Technology, Changsha 410073, China}

\author{Man-Chao Zhang}
\affiliation{Department of Physics, College of Liberal Arts and Sciences, National University of Defense Technology, Changsha 410073, China}
\affiliation{Interdisciplinary Center for Quantum Information, National University of Defense Technology, Changsha 410073, China}

\author{Yi Xie}
\affiliation{Department of Physics, College of Liberal Arts and Sciences, National University of Defense Technology, Changsha 410073, China}
\affiliation{Interdisciplinary Center for Quantum Information, National University of Defense Technology, Changsha 410073, China}

\author{Chun-Wang Wu}
\affiliation{Department of Physics, College of Liberal Arts and Sciences, National University of Defense Technology, Changsha 410073, China}
\affiliation{Interdisciplinary Center for Quantum Information, National University of Defense Technology, Changsha 410073, China}

\author{Ting Chen}
\affiliation{Department of Physics, College of Liberal Arts and Sciences, National University of Defense Technology, Changsha 410073, China}
\affiliation{Interdisciplinary Center for Quantum Information, National University of Defense Technology, Changsha 410073, China}

\author{Bao-Quan Ou}
\affiliation{Department of Physics, College of Liberal Arts and Sciences, National University of Defense Technology, Changsha 410073, China}
\affiliation{Interdisciplinary Center for Quantum Information, National University of Defense Technology, Changsha 410073, China}

\author{Wei Wu}
\affiliation{Department of Physics, College of Liberal Arts and Sciences, National University of Defense Technology, Changsha 410073, China}
\affiliation{Interdisciplinary Center for Quantum Information, National University of Defense Technology, Changsha 410073, China}

\author{Hui Jing}
\email{jinghui73@foxmail.com}
\affiliation{Key Laboratory of Low-Dimensional Quantum Structures and
	Quantum Control of Ministry of Education, and Department of Physics,
	Hunan Normal University, Changsha 410081, China}

\author{Ping-Xing Chen}
\email{pxchen@nudt.edu.cn}
\affiliation{Department of Physics, College of Liberal Arts and Sciences, National University of Defense Technology, Changsha 410073, China}
\affiliation{Interdisciplinary Center for Quantum Information, National University of Defense Technology, Changsha 410073, China}

\date{\today}

\begin{abstract}
Parity-time($\mathcal{PT}$)-symmetric systems, featuring real eigenvalues despite its non-Hermitian nature, have been widely utilized to achieve exotic functionalities in the classical realm, such as loss-induced transparency or lasing revival. By approaching the exceptional point (EP) or the coalescences of both eigenvalues and eigenstates, unconventional effects are also expected to emerge in pure quantum $\mathcal{PT}$ devices. Here, we report experimental evidences of spontaneous $\mathcal{PT}$ symmetry breaking in a single cold $^{40}\mathrm{Ca}^{+}$ ion, and more importantly, a counterintuitive effect of perfect quantum coherence occurring at the EP. Excellent agreement between experimental results and theoretical predictions is identified. In view of the versatile role of cold ions in building quantum memory or processor, our experiment provides a new platform to explore and utilize pure quantum EP effects, with diverse applications in quantum engineering of trapped ions.
\end{abstract}


\maketitle

In conventional quantum mechanics, Hermiticity is a fundamental axiom ensuring real eigenvalues of physical observables~\cite{001}. A striking discovery in recent years has revealed parity-time($\mathcal{PT}$)-symmetric Hamiltonians~\cite{PTReview1,PTReview2,PTReview3}, despite of their non-Hermitian nature, can also have real eigenvalues \cite{002,003}. By continuously tuning parameter values, spontaneous $\mathcal{PT}$ symmetry breaking can occur at an exceptional point (EP)~\cite{EPReview,S1}, where both the eigenvalues and the eigenstates of the system coalesce. As a result, many counterintuitive phenomena \cite{005,006,007,008,009,010,Non-Hermitian} emerge in such systems, e.g., single-mode lasing or anti-lasing~\cite{SML1,SML2}, loss-induced transparency or lasing~\cite{025,LIL}, EP-enhanced sensing~\cite{EPSensing1,EPSensing2}, to name only a few. These seminal experiments, however, have been performed mainly in the classical realm, and more exotic effects are expected to occur in pure quantum $\mathcal{PT}$ devices.

Achieving $\mathcal{PT}$ symmetry, in principle, requires an exact balance of gain and loss, which is challenging in quantum realm, since practical systems can be unstable in the presence of gain-amplified noises~\cite{024,PTQ}. To overcome this obstacle, passive devices with hidden $\mathcal{PT}$ symmetry were proposed by coupling Hermitian systems to a dissipative reservoir~\cite{LIL,025}. The emergence of EPs in such lossy devices, without any active gain, has been demonstrated in very recent experiments using optical or solid-state systems~\cite{QW,Tomo}, opening up a practical route to observe and manipulate quantum EP effects~\cite{QW,Tomo}. For an example, quantum coherence protection was observed in a $\mathcal{PT}$-broken superconducting circuit~\cite{Tomo}, with post-selection of the experimental results and an exponentially decreasing success rate for longer times. These pioneering experiments on quantum EP systems~\cite{024,QW,Tomo} have provided the important first steps towards emerging non-Hermitian quantum technologies.

In this Letter we report the first experiment on spontaneous breaking of $\mathcal{PT}$ symmetry occurring in a single $^{40}\mathrm{Ca}^{+}$ ion. We note that trapped cold ions having a coherence time as long as 10 minutes \cite{cohtime} have been widely used for quantum memory \cite{memory1, memory2, memory3}, quantum state preparation \cite{statepre}, quantum simulation \cite{simulation1,simulation2,simulation3,simulation4}, high-precision metrology \cite{metrology1, metrology2}, and is viewed as a powerful candidate for building quantum computers. However, up to date, experimental realization of $\mathcal{PT}$ symmetry in such a typical system has not been achieved, hindering its applications in non-Hermitian quantum control of ions. Here we fill this gap by demonstrating clear signatures of quantum EP in a $^{40}\mathrm{Ca}^{+}$ ion. We deterministically demonstrate EP features by measuring the ion-state populations both in the $\mathcal{PT}$-symmetric {(PTS)} regime and in the $\mathcal{PT}$-broken {(PTB)} regime. Furthermore, we observe a tuning point of the non-diagonal elements of the density matrix by approaching the EP, due to which giant enhancement of quantum coherence \cite{quantumcoherence} can be achieved for the system. Our work, as the first demonstration of single-ion $\mathcal{PT}$ symmetry breaking, provides a new platform to explore and utilize more truly quantum EP effects with applications in quantum engineering of trapped ions.


\begin{figure*}[ht]
	\centering
	{
		\includegraphics[width= 0.98\textwidth]{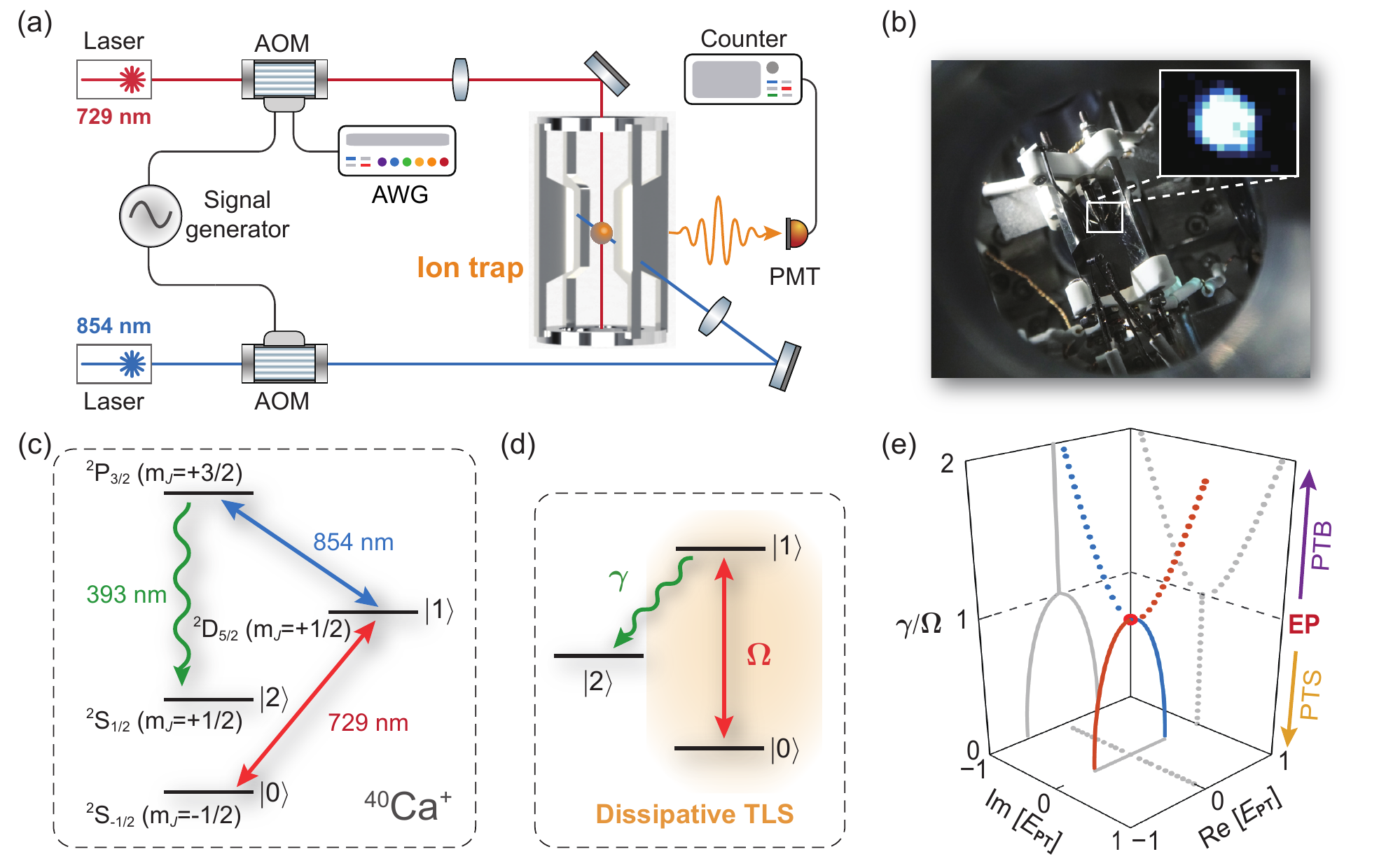}
	}
	\caption{\label{Fig1} The experimental system of passive $\mathcal{PT}$-symmetric single ion. (a) Schematic diagram of the experimental setup. In the experiment, both 729 nm and 854 nm laser beams are switched on at the same time, and the quantum states of the ion at different times are read out by the electron shelving. Here, AOM denotes the acousto-optical modulator, PMT is the photomultiplier tube, and AWG is the arbitrary waveform generator. (b) The photograph of the ion trap. (c-d)  The energy levels of the $^{40}\mathrm{Ca}^{+}$ ion, with the internal states $|0\rangle$, $|1\rangle$ and $|2\rangle$ corresponding to the energy levels $^{2}\mathrm{S}_{1/2}(m_{J}=-1/2)$, $^{2}\mathrm{D}_{5/2}(m_{J}=+1/2)$ and $^{2}\mathrm{S}_{1/2}(m_{J}=+1/2)$, respectively. (e) The eigenvalues of $H_{\mathcal{PT}}$ versus $\gamma/\Omega$. The projection on the back (dotted lines) and the left (thick lines) show the evolution of the imaginary parts and real parts, respectively. The projection on the bottom shows the evolution of the eigenfrequencies in the complex plane, and the EP corresponds to $\gamma/\Omega=1$.}
\end{figure*}

\begin{figure*}[htbp]
\includegraphics[width= 0.99\textwidth]{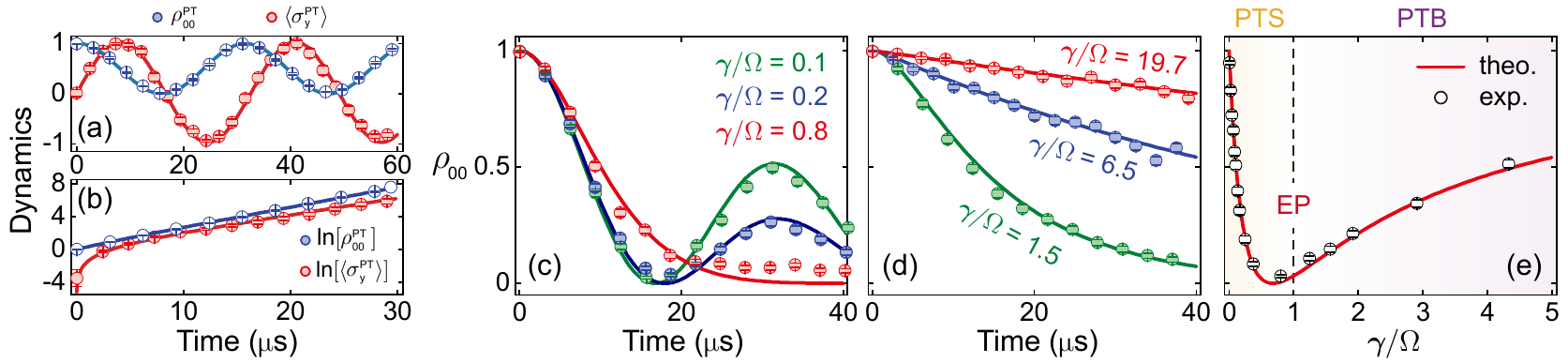}
\caption{\label{Fig2} (a-b) Dynamics of the $\mathcal{PT}$-symmetric system in the initial state $|0\rangle$ in the PTS phase (a) and PTB phase (b), respectively. The parameters are $\Omega = 2\pi\times 32 \ \mathrm{kHz}$, $\gamma = 2\pi\times 1 \ \mathrm{kHz}$ (a), $\gamma = 2\pi\times 47\ \mathrm{kHz}$ (b). (c-d) The dynamics of $\rho_{00}(t)$ of the experimental TLS for two phases show oscillatory (c) to steady state behaviour (d). (e) $\rho_{00}$ at a fixed time $t=2 \pi/\Omega$ versus the loss rate $\gamma$ with $\Omega=2\pi\times 32\ \mathrm{kHz}$. The circles or marks are the experimental data, while the lines are from the theoretical fits. The error bars are the standard deviation of the measurements.}
\end{figure*}

The experimental setup of the trapped $^{40}\mathrm{Ca}^{+}$ ion, with its energy levels, is shown in Figs.~\ref{Fig1}(a-c). The ion, initially prepared in the ground state $|0\rangle = |^{2}\mathrm{S}_{1/2}(m_{J}=-1/2)\rangle$, {is} driven to the excited state $|1\rangle = |^{2}\mathrm{D}_{5/2}(m_{J}=+1/2)\rangle$ by a laser at wavelength 729\,nm. Another laser at 854\,nm induces a tunable loss $\gamma$ in $|1\rangle$, by coupling $|1\rangle$ to a short-life level $|^2\mathrm{P}_{3/2}(m_{J}=+3/2)\rangle$ (which decays quickly to the state $|2\rangle = |^{2}\mathrm{S}_{1/2}(m_{J}=+1/2)\rangle$). This configuration allows the system exhibit coherent transition between $|0\rangle$ and $|1\rangle$, with $|1\rangle$ experiencing the required tunable loss, see {Fig.~\ref{Fig1}(d)}. The effective two-level system {(TLS)}, with coherent transition and tunable loss, is well described by the non-Hermitian Hamiltonian
\begin{eqnarray}
H_{\mathrm{eff}} = \frac{\Omega}{2} \sigma_x - i \gamma |1\rangle\langle 1| \equiv H_{\mathcal{PT}}-i\frac{\gamma}{2}\mathbf{I},\label{eq1}
\end{eqnarray}
where $H_{\mathcal{PT}} = \frac{\Omega}{2} \sigma_x - i\frac{\gamma}{2} \sigma_z$ is the $\mathcal{PT}$-symmetric Hamiltonian with balanced gain and loss,  $\sigma_{x(z)}$ is the Pauli matrix, and $\mathbf{I}$ is the identity operator. The spontaneous $\mathcal{PT}$ symmetry breaking in such a system then arises due to the interplay of the gain-loss rate ($\gamma/2$) and the coupling rate $\Omega/2$~\cite{EPReview}. As shown in {Fig.~\ref{Fig1}(e)}, when the gain-loss rate is smaller than the coupling rate between the two state $(\gamma /\Omega < 1)$, the system exhibits a real spectrum and simultaneous eigenmodes of the Hamiltonian associated with oscillatory solutions. This region is referred to as PTS phase. Yet when the gain-loss rate is bigger than the coupling rate ($\gamma / \Omega >1$), we call it PTB phase where complex conjugate eigenvalues emerge, and one of the eigenmodes exponentially grows \cite{supplement}. The transition between the PTS and PTB phases takes place at an EP which emerges for $\gamma = \Omega$ \cite{EPReview,S1}.


Now we examine these predictions in our experiment. We first verify the dynamical features of this system at different phases.  We initialize the system in $|0\rangle$ and tune the coupling rate $\Omega = 2\pi\times 32\ \mathrm{kHz}$ at time $t =0$, and the value of loss rate can be well controlled. We characterize the $\mathcal{PT}$ symmetry-breaking transitions {by} using the populations of $|0\rangle$ and the coherence in the {$\{|1\rangle, |0\rangle\}$} qubit manifold, which have the following forms \cite{supplement}
\begin{eqnarray}
\rho^{\mathcal{PT}}_{00} (t) &=& \left[\left(2 \gamma^2 - \Omega^2\right) \cosh\left(
	\kappa t \right)\right. \nonumber \\
	&& \left.+ 2 \gamma \kappa \sinh\left(\kappa t \right)-\Omega^2 \right] / 2\kappa^2, \\
\langle \sigma^{\mathcal{PT}}_y (t)\rangle &=& \mathrm{Tr}\left[\sigma_y \rho^{\mathcal{PT}}\right]  \nonumber \\
&=&  {\Omega} \left[-\gamma + \gamma \cosh(\kappa t) + \kappa \sinh(\kappa t)\right]/{\kappa^2},
\end{eqnarray}
with $\kappa = \sqrt{\gamma^2 - \Omega^2}$.  We see that when $\gamma > \Omega$, $\kappa$ is real and the system evolves as $e^{-\kappa t}$ and $e^{\kappa t}$ (when $\kappa t \gg 1$, just $e^{\kappa t}$ remains) \cite{supplement}. But when $\gamma < \Omega$, $\sinh(\kappa t)$ or $\cosh(\kappa t)$ corresponds to the time evolution $e^{\pm i|\kappa|t}$, featuring oscillatory evolution at angular frequency $|\kappa|$ \cite{S2}.

In our experiment, directly measured quantities are the density matrix elements of $\rho(t)$. We get the experiment data of $\rho^{\mathcal{PT}}$ from the relation $\rho^{\mathcal{PT}}(t)=e^{\gamma t}\rho(t)$, which can be easily derived from Eq.\,(\ref{eq1}). The experimental results are shown in {Figs.~\ref{Fig2}(a-b)}. When $\gamma / \Omega$ is tuned to the PTS phase by varying the laser power, the population and the coherence exhibit oscillation with frequency $|\kappa|$, while in the PTB phase, both of the population $\rho^{\mathcal{PT}}_{00} $ and the coherence $\langle \sigma^{\mathcal{PT}}_y (t)\rangle $ increase exponentially. All the experimental results agree well with the theoretical phase diagram \cite{supplement}.


The $\mathcal{PT}$-symmetry-breaking phase transition can also be verified in our the lossy TLS. As shown in Figs.~\ref{Fig2}(c-d), dynamical behaviors of the TLS are clearly different in PTS and PTB phases. The  population of state $|0\rangle$ features decaying oscillations in the PTS phase, and meanwhile, the evolution is accelerated with the increase of loss rate $\gamma$ ({Fig.~\ref{Fig2}(c)}). In contrast, in the PTB phase, the population of state $|0\rangle$ monotonically decays during the system evolving to a steady state, and the evolution is slowed down with the increase of loss rate $\gamma$ (Fig.~\ref{Fig2}(d)). This suggests a possible relation between $\mathcal{PT}$ symmetry and quantum Zeno effect\,\cite{anti-zeno,zeno1,zeno2}, as proposed very recently\,\cite{QZ/AZ01,QZ/AZ02,026}, when the projection measurement of $|1\rangle$ is induced by extremely strong loss. This new possibility will be investigated elsewhere.

We also measure the population of $|0\rangle$ at different loss rates for a fixed time $t = 2\pi/\Omega$, with $\Omega = 2\pi\times 32\ \mathrm{kHz}$, as shown in {Fig.~\ref{Fig2}(e). Clearly, a turning point is observed for the population of $|0\rangle$, which can be well explained by considering the EP of {$H_{\mathrm{eff}}$}: when the loss rate $\gamma=0$, the populations of the two states $|0\rangle$ and $|1\rangle$ can be exchanged freely with each other; by increasing the loss rate ($\gamma<\Omega$),  the state $|1\rangle$ decays to $|2\rangle$ faster via the dissipative channel and hence the population of $|0\rangle$ also decreases faster. In particular, at the EP ($\gamma=\Omega$), the coherent coupling balances with the loss of $|1\rangle$ state, thus the population of $|0\rangle$ reaches its minimum. After this point ($\gamma > \Omega$), as the loss increased, the $|0\rangle$ state will become localized, which can be explained by quantum Zeno effect \cite{026}. The observed phenomenon is similar to the loss-induced lasing reported in a classical system\,\cite{025}. We note that $\gamma_{\textup{min}}$ in Fig.~\ref{Fig2}(e)} is not exactly the EP, since the system is not at the final steady state when we do the measurement at time $t=2\pi/\Omega$. By setting the time long enough, we can have $\gamma_{\textup{min}}$ being closer to the EP (see also Ref.\,\cite{025}).

In order to further visualize pure quantum features of the phase transition, we now introduce the order parameter by the time average of $\langle\sigma_z\rangle$ as defined in Ref.\,\cite{supplement,027, S3}
\begin{eqnarray}\label{sigmaZ}
\Sigma_{Z}(\gamma)&=&\lim_{T\rightarrow\infty}\frac{1}{T}\int_{0}^{T} \frac{\langle \sigma_{z}(\gamma,t)\rangle}{\mathrm{Tr}[\rho (\gamma,t)]}  \mathrm{d}t \nonumber \\ &=& \left\{
\begin{aligned}
 & 0, &&0<\gamma < \Omega,  \\
 & -\sqrt{\gamma^2 - \Omega^2}/{\gamma}, &&\gamma \ge \Omega,
\end{aligned}
\right.
\end{eqnarray}
and the order parameter by the time average of $\langle\sigma_y\rangle$ \cite{supplement}
\begin{eqnarray}\label{sigmaY}
\Sigma_{Y}(\gamma)&=&\lim_{T\rightarrow\infty}\frac{1}{T}\int_{0}^{T}  \left| \frac{\langle \sigma_{y}(\gamma,t)\rangle}{\mathrm{Tr}[\rho(\gamma,t)]} \right|\mathrm{d}t \nonumber \\ &=& \left\{
\begin{aligned}
 & {\Omega}\left[\pi-{2 \arccos({\gamma}/{\Omega})}\right]/({\gamma}{\pi}), &&0< \gamma < \Omega,  \\
 & {\Omega}/{\gamma},  &&\gamma \ge \Omega.
\end{aligned}
\right.
\end{eqnarray}

\begin{figure}[htbp]
	\includegraphics[width= 0.48\textwidth]{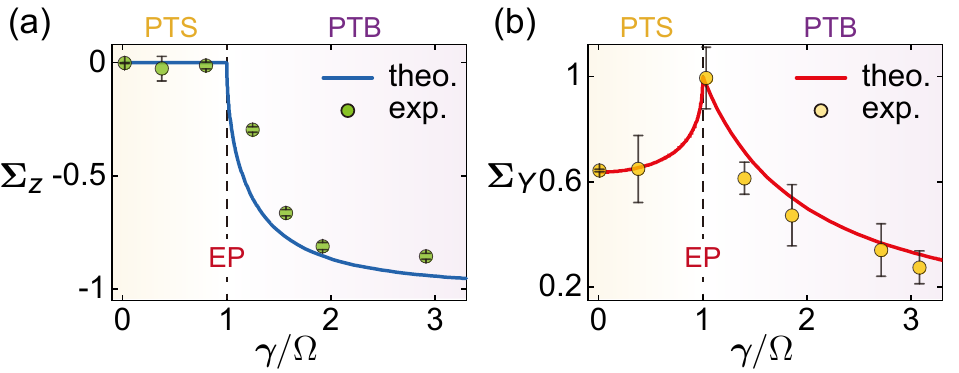}
	\caption{\label{Fig4} The order parameters  $\Sigma_{Z}$ (a) and $\Sigma_{Y}$ (b) versus $\gamma$, with $\Omega = 2\pi\times 32\  \mathrm{kHz}$ and the initial state $|0\rangle$. The error bars indicate the estimated error from the fit to the dynamical equation for $\rho(t)$. }
\end{figure}
We note that for open systems as the experimental TLS and the $\mathcal{PT}$ system, the traces of $\rho(t)$ and $\rho^{\mathcal{PT}}(t)$ are in general not conserved, and thus renormalization is required to study non-Hermitian dynamics\,\cite{Tomo, 027, 024, 026, S3}. Nevertheless, we have $$\rho^{\mathcal{PT}}/\mathrm{Tr}[\rho^{\mathcal{PT}}]=e^{-\gamma t}\rho/\mathrm{Tr}[e^{-\gamma t}\rho]=\rho/\mathrm{Tr}[\rho],$$ hence the dynamical behaviors of $\rho(t)$ and $\rho^{\mathcal{PT}}(t)$ remain the same after the renormalization process. The order parameter $\Sigma_{Z(Y)}$, determined by $\gamma /\Omega$, is independent on the initial state and clearly exhibits the transition feature for $\gamma/\Omega=1$.
We also see that at the EP, the populations of $|0\rangle$ and $|1\rangle$ are equal, i.e., the system stays in a coherent superposition state $ (|0\rangle-i|1\rangle)/\sqrt{2}$ \cite{supplement}.

Fig.~\ref{Fig4}(a) shows the experimental results about $\Sigma_{Z}$ or the mean energy of the system. In the PTS phase, the populations of $|0\rangle$ and $|1\rangle$ are the same due to their periodic oscillations, thus leading to zero for $\Sigma_{Z}$; in contrast, the system can reach the steady state in the PTB phase. These results can also be observed in classical $\mathcal{PT}$ systems, since $\Sigma_{Z}$ can only show the mean energy. In order to see true quantum $\mathcal{PT}$ features, it is necessary to study the other order parameter of the system, such as $\Sigma_Y$.

Fig.~\ref{Fig4}(b) shows quantum coherence $\Sigma_{Y}$  of the $\mathcal{PT}$-symmetric single-ion system, quantified by the modulation of non-diagonal elements of the normalized density matrix. We stress that this is the first example on experimental observations of the order parameter $\Sigma_{Y}$, which was not given in recent quantum $\mathcal{PT}$ experiments\,\cite{024,026}. We find that $\Sigma_{Y}$ reaches its maximum when approaching the EP, i.e., quantum coherence of the $\mathcal{PT}$ system can be enhanced at the EP. This is because at the EP, the balance between the loss and the coherent transition can be reached and thus enables the system to stay in coherent superposition of $|0\rangle$ and $|1\rangle$. We also note that very recently, quantum discord enhancement was observed in an anti-$\mathcal{PT}$-symmetric atomic system by engineering dissipative coupling of optical channels~\cite{APT}.

In conclusion, we have observed spontaneous $\mathcal{PT}$ symmetry breaking in a single $^{40}\mathrm{Ca}^{+}$ ion. We find that when the system is steered to an EP and past it, both the mean populations of the ion states and the quantum coherence exhibit a turning point. To the best of our knowledge, this is the first work on experimental observation of $\mathcal{PT}$ symmetry in a single trapped ion, revealing the counterintuitive EP-enabled effect of quantum coherence enhancement (see also Ref.\,\cite{APT}). In view of the long coherence time of trapped ions, together with well developed techniques of engineering their quantum states, our work provides a powerful new tool for exploring and utilizing true quantum EP effects at single-ion levels. In a broader view, our work can also help to design and utilize unconventional ion-based devices, such as non-Hermitian quantum memory and EP-enhanced quantum processor or cold-ion EP clock.

The authors thank Ran Huang for helpful discussions. This work is supported by the National Basic Research Program of China under Grant No. 2016YFA0301903, the National Science Foundation of China under Grant No.\,11935006, No. 12074433, No. 61632021, No.\,11774086, No.11871472 and the Natural Science Foundation of Hunan Province of China under Grant No. 2018JJ2467.

\providecommand{\noopsort}[1]{}\providecommand{\singleletter}[1]{#1}%
%


\end{document}


\appendix
\setcounter{equation}{0}
\renewcommand{\theequation}{S\arabic{equation}}
\setcounter{figure}{0}
\renewcommand{\thefigure}{S\arabic{figure}}

\title{Supplemental material:
Observation of $\mathcal{PT}$-symmetric quantum coherence in a single ion system}
	
\author{Wei-Chen Wang}
\thanks{These authors contributed equally to this work.}
\affiliation{Department of Physics, College of Liberal Arts and Sciences, National University of Defense Technology, Changsha 410073, China}
\affiliation{Interdisciplinary Center for Quantum Information, National University of Defense Technology, Changsha 410073, China}

\author{Yan-Li Zhou}
\thanks{These authors contributed equally to this work.}
\affiliation{Department of Physics, College of Liberal Arts and Sciences, National University of Defense Technology, Changsha 410073, China}
\affiliation{Interdisciplinary Center for Quantum Information, National University of Defense Technology, Changsha 410073, China}

\author{Hui-Lai Zhang}
\affiliation{Key Laboratory of Low-Dimensional Quantum Structures and
	Quantum Control of Ministry of Education, and Department of Physics,
	Hunan Normal University, Changsha 410081, China}

\author{Jie Zhang}
\affiliation{Department of Physics, College of Liberal Arts and Sciences, National University of Defense Technology, Changsha 410073, China}
\affiliation{Interdisciplinary Center for Quantum Information, National University of Defense Technology, Changsha 410073, China}

\author{Man-Chao Zhang}
\affiliation{Department of Physics, College of Liberal Arts and Sciences, National University of Defense Technology, Changsha 410073, China}
\affiliation{Interdisciplinary Center for Quantum Information, National University of Defense Technology, Changsha 410073, China}

\author{Yi Xie}
\affiliation{Department of Physics, College of Liberal Arts and Sciences, National University of Defense Technology, Changsha 410073, China}
\affiliation{Interdisciplinary Center for Quantum Information, National University of Defense Technology, Changsha 410073, China}

\author{Chun-Wang Wu}
\affiliation{Department of Physics, College of Liberal Arts and Sciences, National University of Defense Technology, Changsha 410073, China}
\affiliation{Interdisciplinary Center for Quantum Information, National University of Defense Technology, Changsha 410073, China}

\author{Ting Chen}
\affiliation{Department of Physics, College of Liberal Arts and Sciences, National University of Defense Technology, Changsha 410073, China}
\affiliation{Interdisciplinary Center for Quantum Information, National University of Defense Technology, Changsha 410073, China}

\author{Bao-Quan Ou}
\affiliation{Department of Physics, College of Liberal Arts and Sciences, National University of Defense Technology, Changsha 410073, China}
\affiliation{Interdisciplinary Center for Quantum Information, National University of Defense Technology, Changsha 410073, China}

\author{Wei Wu}
\affiliation{Department of Physics, College of Liberal Arts and Sciences, National University of Defense Technology, Changsha 410073, China}
\affiliation{Interdisciplinary Center for Quantum Information, National University of Defense Technology, Changsha 410073, China}

\author{Hui Jing}
\email{jinghui73@foxmail.com}
\affiliation{Key Laboratory of Low-Dimensional Quantum Structures and
	Quantum Control of Ministry of Education, and Department of Physics,
	Hunan Normal University, Changsha 410081, China}

\author{Ping-Xing Chen}
\email{pxchen@nudt.edu.cn}
\affiliation{Department of Physics, College of Liberal Arts and Sciences, National University of Defense Technology, Changsha 410073, China}
\affiliation{Interdisciplinary Center for Quantum Information, National University of Defense Technology, Changsha 410073, China}

\maketitle

\section{S1: The analytical form of the dynamics of the system}

\subsection{A. Density matrix form of $\rho(t)$ and $\rho^{\mathcal{PT}}$}
This full system can be considered as a three-level system, where $|0\rangle \leftrightarrow |1\rangle$ transition is driven by the laser with Rabi frequency $\Omega$, and the decay $|1\rangle \rightarrow |2\rangle$ corresponds the loss of the system. The time evolution of the full system obeys a Lindblad master equation  $\dot{\varrho}(t) = \mathbf{L}\varrho  $, where the generator of the dynamics $\mathbf{L}$ is normally called Liouvillian superoperator, and has the form
\begin{eqnarray}
\mathbf{L}( \cdot) : = - i [H, (\cdot)] + \big(J(\cdot)J^{\dagger} - 2\{J^{\dagger}J,\ (\cdot)\} \big).
\end{eqnarray}
Here $\varrho$ is the density matrix of the three-level system, $H = \frac{\Omega}{2}(|0\rangle \langle 1| + |1\rangle \langle 0|)$ is the Hamiltonian and $J = \sqrt \gamma |2\rangle \langle 1|$ is the quantum jump operators, with $\gamma$ the tunable decay rate. From equation (S1), we get
\begin{subequations}\label{subequation}
	\begin{align}
	\dot{\varrho}_{11} =&\ -2\gamma \varrho_{11} - i \frac{\Omega}{2} (\varrho_{01} - \varrho_{10}),\label{sub1}\\
	\dot{\varrho}_{00} =&\ i \frac{\Omega}{2} (\varrho_{01} - \varrho_{10}),\label{sub2}\\
	\dot{\varrho}_{01} =&\ -\gamma \varrho_{01} + i\frac{\Omega}{2} (\varrho_{00} - \varrho_{11}).\label{sub3}
	\end{align}
\end{subequations}
So the density matrix $\rho(t)$ for the dissipative TLS is determined by a lower dimension  Liouvillian superoperator $\dot{\rho} = \mathcal L \rho(t)$ with
\begin{equation}
\mathcal L = \begin{pmatrix}
-2\gamma & i\frac{\Omega}{2} & -i\frac{\Omega}{2} & 0\\
i\frac{\Omega}{2} & -\gamma & 0 & -i\frac{\Omega}{2}\\
-i\frac{\Omega}{2} & 0 & -\gamma & i\frac{\Omega}{2}\\
0 & -i\frac{\Omega}{2} & i\frac{\Omega}{2} & 0
\end{pmatrix}.
\end{equation}
If we define an effective, non-Hermitian Hamiltonian by
$H_{\mathrm{eff}} = \frac{\Omega}{2}\sigma_x - i\gamma |1\rangle \langle 1|$, where $\sigma_x$ is the Pauli matrix, we can rewrite the master equation of TLS like
\begin{eqnarray}
\dot \rho(t) = \mathcal L \rho = -i [H_{\mathrm{eff}}\rho - \rho H_{\mathrm{eff}}^{\dagger}],
\end{eqnarray}
where we have introduced the Liouvillian superoperator $\mathcal L$ without quantum jumps. This means that Liouvillian superoperator $\mathcal L$ plays the same role with the non-Hermitian Hamiltonian. Since the Liouvillian $\mathcal L $ is a non-Hermitian matrix, it too can exhibit EPs, which can be defined as the degeneracy points of $\mathcal L$~\cite{S1}.

\begin{figure}[h]
	\includegraphics[scale=0.88]{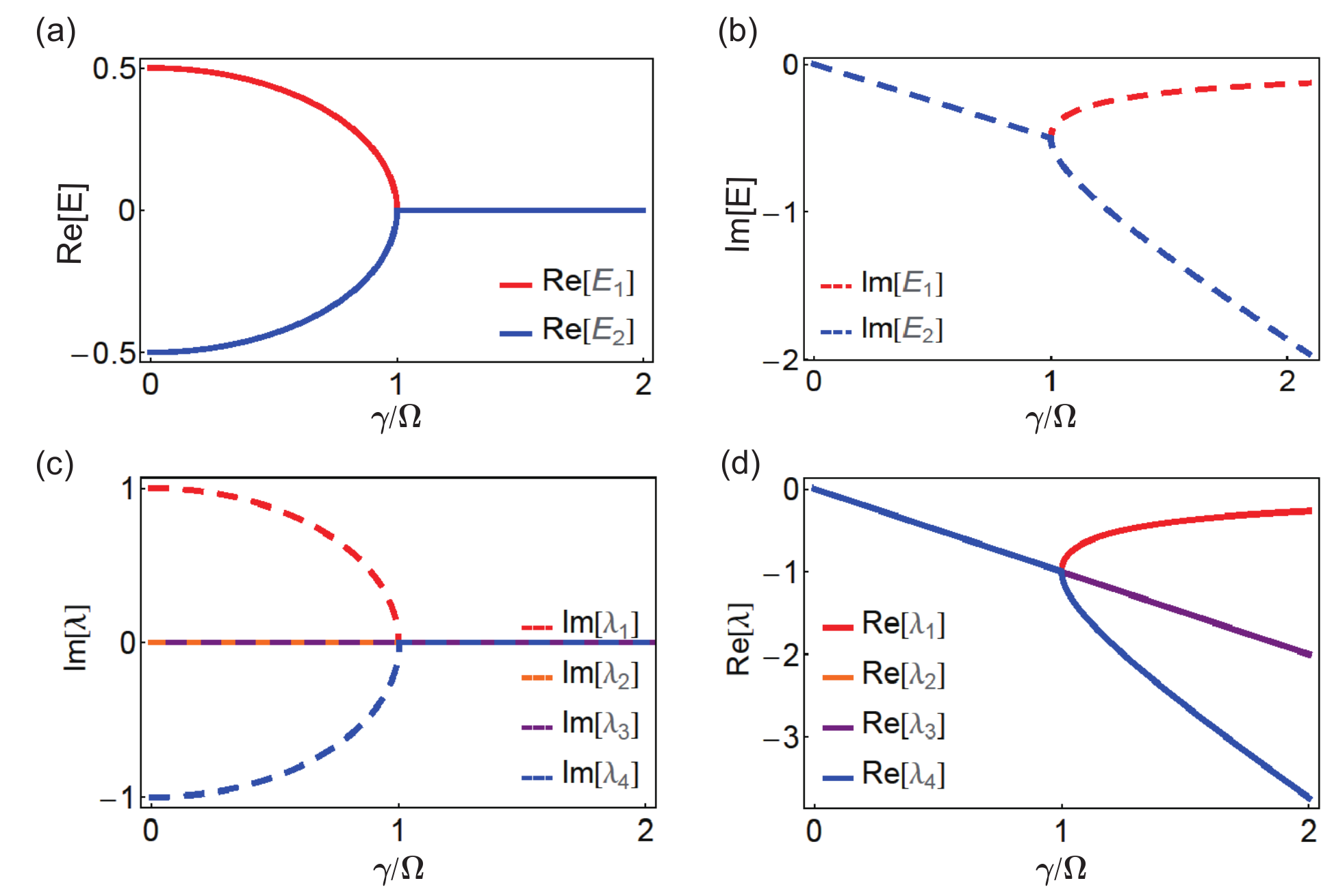}
	\caption{(a) The real parts of the eigenvalues of effective Hamiltonian.  (b) The imaginary parts of the eigenvalues of effective Hamiltonian. (c) The imaginary parts of the eigenvalues of $\mathcal L$. (d) The real parts of the eigenvalues of $\mathcal L$. They show the same EP with $H_{\mathcal{PT}}$ as shown in Fig. 1(d).}
\end{figure}

We get the the eigenvalues and eigenvectors of $H_{\mathrm{eff}}$ via the relation $H_{\mathrm{eff}} |\phi_i\rangle = E_{i} |\phi_i\rangle$ ~\cite{S1}
\begin{subequations}\label{subequation}
	\begin{align}
	\{E\} &= \frac{1}{2}\{ -i\gamma + \sqrt{\Omega^2 - \gamma^2},\  \frac{1}{2}-i\gamma -\sqrt{\Omega^2 - \gamma^2}\}, \label{sub1}\\
	\{\phi\} & \propto \left\{ \{-i\gamma + \sqrt{\Omega^2 - \gamma^2}, \Omega \}^\top, \{-i\gamma - \sqrt{\Omega^2 - \gamma^2}, \Omega \}^\top \right\}.
	\end{align}
\end{subequations}
One can obtain much information about the experimental system by finding the eigenvalues of $\mathcal L$~\cite{S1,S2}. The eigenvalues of  $\mathcal L$ are $\{\lambda_1 = -\gamma + \kappa, \lambda_{2(3)} = -\gamma, \lambda_4 = -\gamma - \kappa \}$ with $\kappa = \sqrt{\gamma^2 - \Omega^2}$ (which we order by decreasing real part, $\mathrm{Re}[\lambda_i] \geq \mathrm{Re}[\lambda_{i+1}]$). Every eigenvalue $\lambda_i$ corresponds to a right eigenmatrix $R_i$ and a left eigenmatrix $L_i$ of $\mathcal L$, which are defined by $\mathcal L R_i = \lambda_i R_i$ and $\mathcal L^{\dagger} L_i = \lambda_i^* L_i$~\cite{S1,S2}. The eigensystem of Liouvillian superoperator $\mathcal L$ can be written as (not normalized)
\begin{subequations}\label{subequation}
	\begin{align}
	\{\lambda\} &= \{ -\gamma + \kappa,\  -\gamma,\  -\gamma,\  -\gamma -\kappa \}, \label{sub1}\\
	\{R\} &\propto \left\{ \begin{pmatrix}
	-\frac{2 \gamma (-\gamma + \kappa) + \Omega^2}{\Omega^2} & i\frac{(-\gamma + \kappa)}{\Omega}\\
	-i\frac{(-\gamma + \kappa)}{\Omega} & 1
	\end{pmatrix},
	\begin{pmatrix}
	1 & -i\frac{\gamma }{\Omega}\\
	0 & 1
	\end{pmatrix},
	\begin{pmatrix}
	0 & 1 \\
	1 & 0
	\end{pmatrix},
	\begin{pmatrix}
	\frac{2 \gamma (\gamma + \kappa) - \Omega^2}{\Omega^2} & -i\frac{(\gamma + \kappa)}{\Omega}\\
	i\frac{(\gamma + \kappa)}{\Omega} & 1
	\end{pmatrix}\right\}.
	\end{align}
\end{subequations}
We can see that the relation between the eigenvalues properties of  $\mathcal L$ and $H_{\mathrm{eff}}$ is $\lambda_i = -i(E_j - E_k^*)$ and the eigenmatrix of $\mathcal L$ is given by $R_i=\{|\phi_j\rangle \langle \phi_k|\}$. Therefore, the EP of $H_{\mathrm{eff}}$ is obviously the same with $\mathcal L $ (see Fig. S1).

At time $t$, the time evolution of the experimental TLS system from the initial state $\rho(0)$ is (except for the EP \cite{S1, S2}) $\rho(t) = \sum_{i=1}^4e^{\lambda_i t}c_i R_i = e^{-\gamma t} \rho^{\mathcal{PT}}(t)$, where
\begin{eqnarray}
\rho^{\mathcal{PT}}(t) = e^{\kappa t}c_1 R_1 + c_2 R_2 + c_3 R_3 + e^{-\kappa t}c_4 R_4
\end{eqnarray}
is the time evolution of $\mathcal{PT}$-symmetric system , $c_i = \mathrm{Tr}[L_i\rho(0)]$ are coefficients of the initial state decomposition onto the eigenbasis of $\mathcal L$ \cite{S2}. If the initial state is $|0\rangle$, we can get the $\mathcal{PT}$ density matrix at time $t$ (not normalized)
\begin{eqnarray}
\rho^{\mathcal{PT}}(t)= \begin{pmatrix}
\frac{\Omega^2 \sinh(\kappa t/2)^2}{\kappa^2} & i\frac{\Omega(\gamma -\gamma \cosh(\kappa t) -\kappa \sinh(\kappa t))}{2\kappa^2} \\
-i\frac{\Omega(\gamma -\gamma \cosh(\kappa t) -\kappa \sinh(\kappa t))}{2\kappa^2} & \frac{-\Omega^2 + (2 \gamma^2 - \Omega^2) \cosh(
	\kappa t) + 2 \gamma \kappa \sinh(\kappa t)}{2\kappa^2}
\end{pmatrix}
\end{eqnarray}
So
\begin{eqnarray}
\langle \sigma^{\mathcal{PT}}_z (t)\rangle &=& -\cosh(\kappa t) - \frac{\gamma \sinh(\kappa t)}{k},\\
\langle \sigma^{\mathcal{PT}}_y (t)\rangle &=& \mathrm{Tr}[\sigma_y \rho^{\mathcal{PT}}] =  \frac{\Omega}{\kappa^2} (-\gamma + \gamma \cosh(\kappa t) + \kappa \sinh(\kappa t))
\end{eqnarray}

\subsection{B. The dynamics after renormalization}

\begin{figure}[h]
	\centering
	{
		\includegraphics[width=16cm]{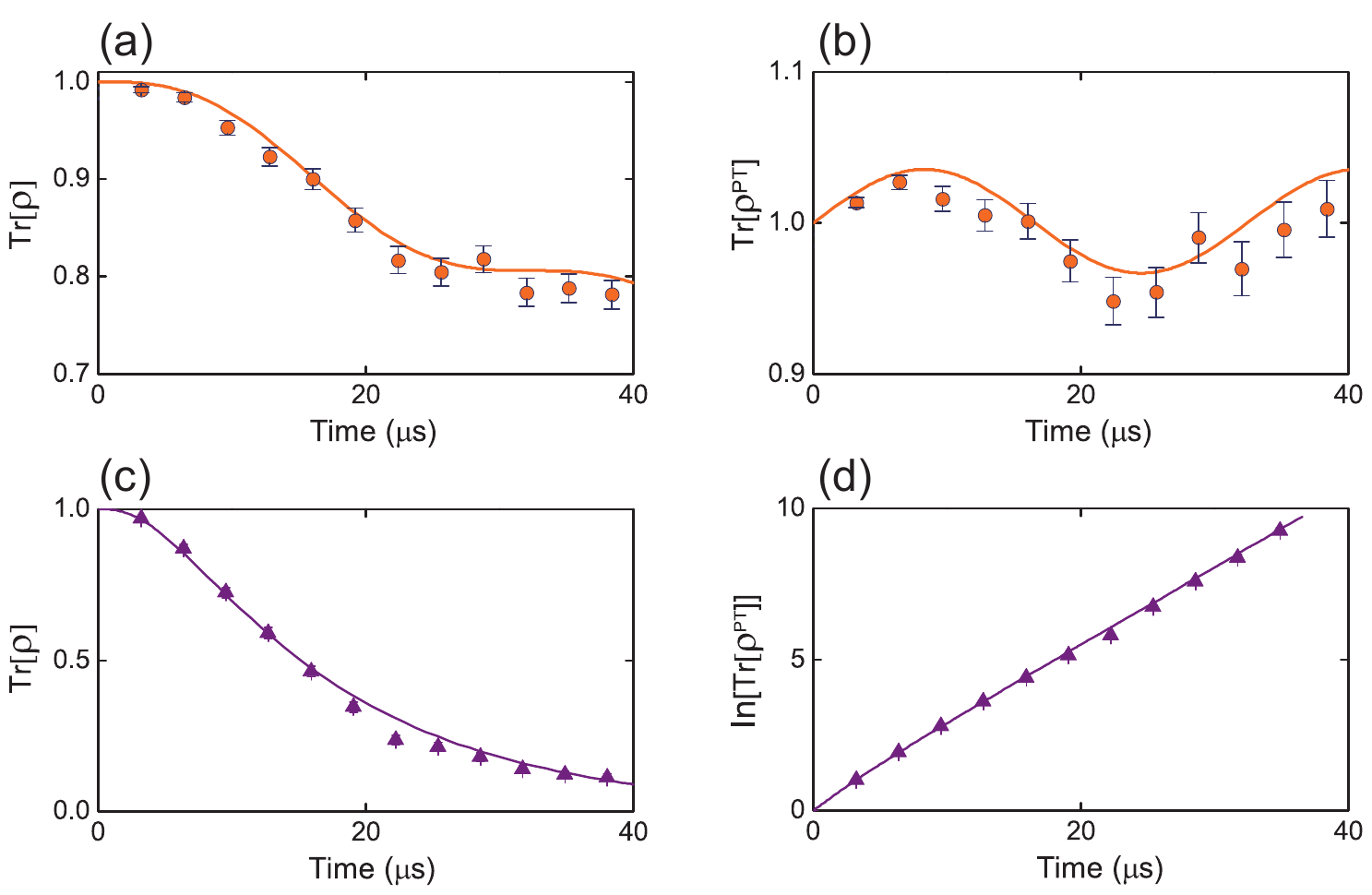}
	}
	\caption{(Color online) (a-b) The trace of the density operator $\rho$ (a) and $\rho^{\mathcal{PT}}$ (b) change with time in the PTS phase, with $\Omega = 2\pi\times 32$ kHz, $\gamma = 2\pi\times 1$ kHz. (c-d) The trace of the density operator $\rho$ (c) and $\rho^{\mathcal{PT}}$ (d) change with time in the PTB phase, with $\Omega = 2\pi\times 32$ kHz, $\gamma =  2\pi\times 47$ kHz.}
\end{figure}

We have to mentioned that both $\rho(t)$ and $\rho^{\mathcal{PT}}(t)$ , on their own, do not preserve the trace of the density operator (see Fig. S2). As shown in Figs. S2(a,c), for $\rho(t)$, which is a dissipative system, the trace is always decay whatever in PTS or PTB phase. But for $\rho^{\mathcal{PT}}(t)$, which has loss and gain, as shown in Figs. S2(b,d), the trace will show $\mathcal{PT}$ symmetry breaking phenomenon. So if we want to know the dynamics of the system, we need renormalize $\rho(t)$ by $\rho (t) = e^{\mathcal L t} \rho(0)/\mathrm{Tr}[e^{\mathcal L t} \rho(0)]$ \cite{S3,S4}. Since $\rho^{\mathcal{PT}}/\mathrm{Tr}[\rho^{\mathcal{PT}}]=e^{-\gamma t}\rho/\mathrm{Tr}[e^{-\gamma t}\rho]=\rho/\mathrm{Tr}[\rho]$, the dynamical behaviors of $\rho(t)$ and $\rho^{\mathcal{PT}}(t)$ remain the same after the renormalization process, which shows the equivalence of the two systems on the PT phase transition.

Eqs. (S7,S8) tell us the dynamical behaviors of the system. (i) When $0<\gamma < \Omega$, $\kappa$ is an imaginary number, $\rho^{\mathcal{PT}}$ will oscillate periodically at frequency $|\kappa|$. When $\gamma$ increase (but still smaller than $\Omega$), the oscillating frequency will get smaller and smaller. (ii) Till $\gamma = \Omega$, which is the EP, not only all the eigenvalues of $\mathcal L$ is degenerated, but also the eigenmatrix $R_1$ and $R_4$ are degenerated, and
\begin{eqnarray}
R_{EP} = R_{1(4)}(\gamma = \Omega) = \frac{1}{2}\begin{pmatrix}
1 & -i\\
i& 1
\end{pmatrix}.
\end{eqnarray}
This means that at EP, the system will be in the state $R_{EP}$ , which is a maximum coherent superposition state $(|0\rangle - i|1\rangle)/\sqrt 2$.
(iii) In the PTB phase, $\kappa > 0 $, when $\kappa t \gg 1$,  $e^{-\kappa t} \rightarrow 0$ and the first mode $R_1$ dominates the dynamics of the system. That is to say $\rho(t)$ will evolve to the eigenmode $R_1$ finally, whatever the initially state is. We characterize this in terms of the trance distance $||\rho(t) - R_1|| = \mathrm{Tr}[\sqrt{(\rho(t) - R_1)^{\dagger}(\rho(t) - R_1)}]$ between $\rho(t)$ and $R_1$ for different initial state \cite{S2,S3}. The trace distance measures the distinguish ability of two quantum states and the results are shown in Fig. S3. It shows that when the initial state is the eigenmode $R_i$, the state of the system will stay there, we call those states fixed points~\cite{S4}. This is because coefficients $\mathrm{Tr}[L_{j\neq i}R_{i}] = 0$, so $\rho(t) = \rho(0) = R_i$. However, for other initial states, the system will evolve to $R_1$ finally.
\begin{figure}[h]
	\centering
	{
		\includegraphics[width=12cm]{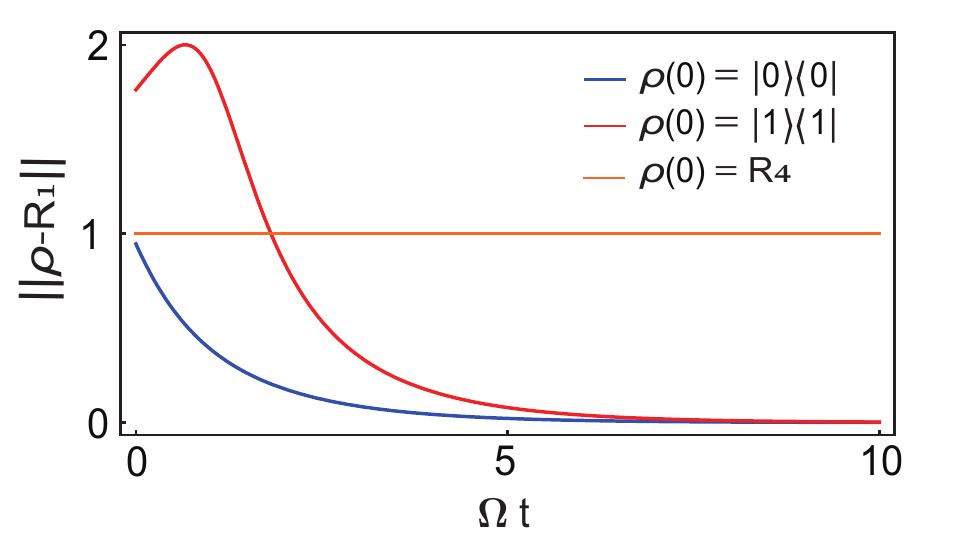}
	}
	\caption{In PTB phase, the system will evolve to the steady state $R_1$, whatever the initial state is, expect the eigenmodes $R_{2,3,4}$. Here we set $\gamma / \Omega = 1.2$. }
\end{figure}

If we define $\langle\widetilde {\sigma}_{z(y)}(t)\rangle=\mathrm{Tr}[\sigma_{z(y)}\rho]/\mathrm{Tr}[\rho]$, we will get
\begin{eqnarray}
\langle\widetilde {\sigma}_{z}(t)\rangle &=& \frac{\kappa (\kappa \cosh(\kappa t) + \gamma \sinh(\kappa t))}{\Omega^2 -\gamma^2 \cosh(\kappa t) -\gamma\kappa \sinh(\kappa t)},\\
\langle\widetilde {\sigma}_{y}(t)\rangle &=& \frac{\Omega (\gamma -\gamma \cosh(\kappa t) - \kappa \sinh(\kappa t))}{\Omega^2 -\gamma^2 \cosh(\kappa t) -\gamma\kappa \sinh(\kappa t)}.
\end{eqnarray}

(i) $\gamma >\Omega$:  The system will evolve to eigenmode $R_1$ finally, so
\begin{eqnarray}
\Sigma_{Z}&=&\lim_{T\rightarrow\infty}\frac{1}{T}\int_{0}^{T} \frac{\mathrm{Tr}[\sigma_{z}\rho]}{\mathrm{Tr}[\rho]} \mathrm{d}t\nonumber \\
&=&\frac{\mathrm{Tr}[\sigma_{z}R_1]}{\mathrm{Tr}[R_1]}\nonumber \\
&=&-\kappa/\gamma
\end{eqnarray}
\begin{eqnarray}
\Sigma_{Y}&=&\lim_{T\rightarrow\infty}\frac{1}{T}\int_{0}^{T}\frac{\mathrm{Tr}[\sigma_{y}\rho]}{\mathrm{Tr}[\rho]} \mathrm{d}t\nonumber \\
&=& \frac{\mathrm{Tr}[\sigma_{y}R_1]}{\mathrm{Tr}[R_1]} \nonumber \\
&=&\Omega/\gamma
\end{eqnarray}

(ii)$\gamma < \Omega$: The evolution of $\langle \widetilde {\sigma}_{z(y)}(t)\rangle$ is periodic, so it is only necessary to calculate the time average of the data within a period. If we define $\omega = \sqrt{\Omega^2-\gamma^2}, \alpha = \arccos(\gamma/\Omega)$, then
\begin{eqnarray}
\langle\widetilde {\sigma}_{z}(t)\rangle &=& \frac{-\omega(\omega \cos(\omega t) + \gamma \sin(\omega t))}{\Omega^2 -\gamma^2 \cos(\omega t) +\gamma\omega \sin(\omega t)},\\
\Sigma_{Z}&=&\int_{0}^{2\pi/\omega} \frac{\langle\widetilde {\sigma}_{z}(t)\rangle}{(2\pi/\omega)} \mathrm{d}t=0
\end{eqnarray}
and
\begin{eqnarray}
\langle\widetilde {\sigma}_{y}(t)\rangle &=& \frac{\Omega}{\gamma}\left(1-\frac{1-(\gamma/\Omega)^2}{1-(\gamma/\Omega)\cos(\omega t + \alpha)}\right),\\
\Sigma_{Y}&=&\int_{0}^{2\pi/\omega} \frac{|\langle\widetilde {\sigma}_{y}(t)\rangle|}{(2\pi/\omega)} \mathrm{d}t\nonumber \\
&=&\frac{1}{2\pi}\left(\int_{\alpha}^{2\pi-\alpha}\langle\widetilde {\sigma}_{y}(x)\rangle \mathrm{d}x +\int_{2\pi-\alpha}^{2\pi+\alpha}-\langle\widetilde {\sigma}_{y}(x)\rangle \mathrm{d}x\right)\nonumber \\
&=&\frac{1}{2\pi}\frac{\Omega}{\gamma}(2\pi-4\alpha)\nonumber \\
&=&\frac{\Omega}{\gamma}\left(1-\frac{2\arccos(\gamma/\Omega)}{\pi}\right).
\end{eqnarray}

\section{S2: Experimental data processing}
In our experimental scheme, a passive $\mathcal{PT}$ symmetric system is constructed by using the internal state of ions, and the energy level structure is shown in main text Figs. 1(c-d). The Zeeman sublevels $^{2}\mathrm{S}_{1/2}(m_{J}=-1/2)$ and $^{2}\mathrm{D}_{5/2}(m_{J}=+1/2)$ of the $^{40}\mathrm{Ca}^{+}$  ion in a magnetic field of 5.2 G  are chosen as system states $|0\rangle$ and $|1\rangle$, and the Zeeman sublevels $^{2}\mathrm{S}_{1/2}(m_{J}=+1/2)$ is considered as an environment. $^{2}\mathrm{D}_{5/2}(m_{J}=+1/2)$ is excited to $^{2}\mathrm{P}_{3/2}(m_{J}=+3/2)$ by the circularly polarized 854 nm laser beam, and according to the selection rule, $^{2}\mathrm{P}_{3/2}(m_{J}=+3/2)$ will only transit to $^{2}\mathrm{S}_{1/2}(m_{J}=+1/2)$ by spontaneous radiation. Hence, the loss rate between the system and the environment can be realized by adjusting the intensity of 854 nm laser beam. Next, we will introduce the experimental data acquisition and data processing.

\subsection{A. Experimental system}
Our experimental system (state  $|1\rangle$ and state  $|0\rangle$) has a loss channel where state  $|1\rangle$ decays to the "environment" (state  $|2\rangle$). So the experimental system is a dissipative system, and the dynamic evolution of the experimental system is not conserved, that is, the population of the system will leak from the state $|1\rangle$ into the "environment". In the ion trap system, the quantum state tomography of single ion internal state can be realized by electron shelving. For example, in main text Figs. 2(c-e), we show the original data collected from the experimental system. 800 times of repeated measurements were performed on each data point to obtain the probability of the system being in state  $|0\rangle$ ($\rho_{00}$), where the error bar of each data point is the measurement error generated by the 800 times of repeated measurements. By using an arbitrary waveform generator (AWG), the relative phase of the laser (729 nm) can be controlled and the quantum state off-diagonal elements can be detected. Finally, the quantum state tomography of single ion internal state can be realized by using the above method.

\subsection{B. Experiment method}
The dissipation of the experimental system can be controlled by the light intensity of 854 nm laser. As the light intensity increases, the loss rate $\gamma$ in the Hamiltonian $H_{\mathrm{eff}}$ increases. In experiment, it is convenient to measure the diagonal elements of the density matrix, so the experimental data of state  $|0\rangle$ is used to fit the loss rate $\gamma$. To ensure the stability of the experimental system, only the light intensity of 854 nm laser is changed to obtain the dynamic evolution data of state  $|0\rangle$ under different light intensity. By solving the master equation of the density matrix, the population of the state  $|0\rangle$ is,
 \begin{eqnarray}
\rho_{00}=\frac{e^{-\gamma t}(-\Omega^{2}+(2\gamma^{2}-\Omega^{2})\cosh(t\sqrt{\gamma^{2}-\Omega^{2}})+2\gamma\sqrt{\gamma^{2}-\Omega^{2}}\sinh(t\sqrt{\gamma^{2}-\Omega^{2}})}{2(\gamma^{2}-\Omega^{2})}.
\end{eqnarray}
We use Eqs. (S20) to fit our experimental data to get $\gamma$. Considering that $\rho^{\mathcal{PT}}=e^{\gamma t}\rho$, we get the data $\rho^{\mathcal{PT}}(t_{0})$ in the $\mathcal{PT}$ symmetric system by multiplying $e^{\gamma t_{0}}$ on the original data $\rho(t_{0})$.

In order to obtain the data in Fig. 3, first of all, the quantum state tomography must be performed for the internal states of ion with different $\gamma$ at any time to obtain a series of original data. Then, the original data is renormalized to obtain $\langle \widetilde {\sigma}_{z(y)}(t)\rangle =\mathrm{Tr}[\sigma_{z(y)}\rho]/\mathrm{Tr}[\rho]$. In theory we need to average $\langle \widetilde {\sigma}_{z(y)}(t)\rangle$ in time from zero to infinity, but from an experimental point of view that's not realistic. It is not difficult to find that when the system is in the PTS regime, the evolution of $\langle \widetilde {\sigma}_{z(y)}(t)\rangle$ is periodic, so it is only necessary to calculate the time average of the data within a period. Experimentally, if $n$ data points are collected within a period $T$ and the time interval is $\delta t=T/n$, then the integration of main text Eqs.(4) is formulated as a summation formula,
 \begin{eqnarray}
 \Sigma_{Z(Y)}=\sum^{n}_{n=1}\frac{ \langle\widetilde {\sigma}_{z(y)}(t_{n})\rangle\delta t}{T}.
\end{eqnarray}
In PTB regime, according to the theory, when the system evolves long enough (i.e. $|\kappa|t \gg 1$), the first eigenmode $R_1$ of Liouvillian superoperator $\mathcal L$ dominates the dynamics of the system. According to our experimental parameters, when $t_e=50\mu s$, $\rho(t)$ is already approaching to $R_1$, so we get $\Sigma_{Z(Y)}\approx \mathrm{Tr}[\sigma_{z(y)}R_1]/\mathrm{Tr}[R_1]$.
